\documentclass[12pt,preprint]{aastex}
\usepackage{amsmath}
\shorttitle{Central engines of III Zw 2 and PG 1407+265}
\shortauthors{Chen, Cao \& Bai}

\slugcomment{Accepted by ApJ}

\begin{document}

\title{The central engines of two unusual radio-intermediate/quiet active galactic nuclei: III Zw 2 and PG 1407+265}

\author{Liang Chen\altaffilmark{1}, Xinwu Cao\altaffilmark{1} and J. M. Bai\altaffilmark{2,3}}

\altaffiltext{1}{Key Laboratory for Research in Galaxies and
Cosmology, Shanghai Astronomical Observatory, Chinese Academy of
Sciences, 80 Nandan Road, Shanghai, 200030, China}
\altaffiltext{2}{National Astronomical Observatories/Yunnan
Observatory, Chinese Academy of Sciences, Kunming, 650011, China}
\altaffiltext{3}{Key Laboratory for the Structure and Evolution of
Celestial Objects, Chinese Academy of Sciences, Kunming, 650011,
China} \altaffiltext{\dag}{E-mail: chenliang@shao.ac.cn (LC);
cxw@shao.ac.cn (XC); baijinming@ynao.ac.cn (JMB)}

\begin{abstract}

We use the accretion disk/corona$+$jet model to fit the multi-band
spectral energy distributions (SEDs) of two unusual
radio-intermediate/quiet quasars. It is found that the optical/UV
emission of III Zw 2 is probably dominated by the emission from the
accretion disk.
The X-ray emission should be dominated by the radiation from the
jet, while the contribution of the disk corona is negligible.
The optical/UV component in the SED of PG 1407+265 can be well
modeled as the emission from the accretion disk, while the IR
component is attributed to the thermal radiation from the dust torus
with an opening angle $\sim 50^\circ$. If the X-ray continuum
emission is dominated by the synchrotron emission of the
jet, the source should be a ``high peak frequency blazar", which
obviously deviates the normal blazar sequence. The observed SED can
also be fitted quite well by the accretion disk/corona model with
the viscosity parameter $\alpha=0.5$. The spectrum of the accretion
disk/corona in PG 1407+265 satisfies the weak line quasar criterion
suggested in Laor \& Davis.

\end{abstract}


\section{Introduction}

In optically selected quasar samples, quasars with similar optical
properties exhibit very different properties in radio bands. The
radio loudness parameter $R$, the ratio of the radio flux at 5~GHz
to the optical flux at B-band ($R\equiv f_{\rm 5GHz}/f_{\rm B}$), is
used as an indicator of the radio properties of quasars
\citep{1989AJ.....98.1195K}. \citet{1989AJ.....98.1195K} found a
dichotomy in the radio loudness distribution for an optically
selected quasar sample. The quasars with radio loudness $R>10$ are
defined as radio-loud (RL) quasars, while those with $R\le 10$ are
named as radio-quiet (RQ) quasars. It was found that RL active
galactic nuclei (AGNs) have powerful relativistic jets, while most
of RQ AGNs have no or very weak jets. The distribution of
radio-loudness may provide useful clues on the jet formation
mechanism of AGNs and the disk-jet connection, which was
explored by many previous authors
\citep[e.g.,][]{2000ApJS..126..133W,
2002AJ....124.2364I,2001ApJ...555..650H,2003MNRAS.346..447C,
2003MNRAS.345.1057M, 2004A&A...414..895F, 2006ApJ...637..669L,
2007ApJ...658..815S, 2011MNRAS.417..184B}. The FIRST (Faint Images
of the Radio Sky at Twenty Centimeters) detected quasars showed that
the radio loudness distribution is not bimodal, but rather
continuous \citep{2000ApJS..126..133W}. It is still debating on
whether the radio-loudness distribution is bimodal or continuous
\citep[see,][]{2002AJ....124.2364I,2003MNRAS.346..447C}.

The radio-loudness parameter can be a good indicator of the
jet properties of AGNs in the sense of statistics.
\citet{1993MNRAS.263..425M} and \citet{1996ApJ...471..106F}
identified a number of quasars with ``radio intermediate loudness".
They suggested that these sources might be relativistically boosted
radio-weak quasars (``radio-weak blazars"). In fact, the very long
baseline interferometry (VLBI) observations have revealed that many
RQ/radio-intermediate AGNs exhibit jet structure and high-brightness
temperature radio cores \citep[e.g.,][]{1998MNRAS.297..366K,
1998MNRAS.299..165B, 2003ApJ...591L.103B, 2005ApJ...621..123U,
2006ApJ...645..856W, 2006A&A...455..161L}. In this work, we choose
two unusual RQ/radio intermediate sources with relativistic jets,
III Zw 2 and PG 1407+265,
to explore the physics of the disk-jet connection in this kind of
the sources. As these two sources have both RL and RQ
characteristics, the present investigation may reveal the relation
of the disk-jet connection with the radio-loudness parameter, which
may provide clues on the origin of the radio-loudness distribution.


\citet{1968ApJ...152.1101A} classified III Zw 2 (PG 0007+106, Mrk
1501, $z=0.089$) as a Seyfert I galaxy, which is also included in
the PG quasar sample \citep{1983ApJ...269..352S}. It is hosted in a
spiral galaxy \citep{1983Natur.303..584H}, which is typical
characteristics of RQ AGNs. The extended radio emission is very weak
compared with its core emission \citep[$\sim 50-100$~mJy at
1.4 GHz,][]{1987MNRAS.228..521U, 2005A&A...435..497B}, and
superluminal motions of the jet components  in this source have been
detected in the VLBA (Very Long Baseline Array) images
\citep{2000A&A...357L..45B}. The apparent velocity of the moving jet
components is $\beta_{\rm app}=1.25\pm0.09$ in units of light speed
is derived from the VLBA images at 43~GHz. III Zw 2
exhibits violent variability throughout all wavebands
\citep[for example, more than an order of magnitude of variability
in radio bands, see][]{2002MNRAS.335..177S}.
\citet{2010RAA....10..707C} modeled the multi-waveband SED of this
sources, and suggested that III Zw 2 is a possible $\gamma$-ray
source and could be detected by the \emph{Fermi}/LAT in the future.
This source possesses typical blazar-like properties, however, it is
included in the radio-intermediate sample
\citep{1996ApJ...471..106F}.

PG 1407+265 is a RQ AGN \citep[$z=0.94$,][]{1989AJ.....98.1195K,
1995ApJ...450..585M} with some unusual properties. Its flux ratios
in the radio, X-ray, and optical wavebands are typical of normal RQ
quasars \citep{2010ApJ...721..562P}. \citet{1995ApJ...450..585M}
found that the emission lines of PG 1407+265 have very small
equivalent widths (except for H$\alpha$), but the full width at
half-maximum (FWHM) reaches $v\approx7000-12000$ km s$^{-1}$. This
is a weak line quasar (WLQ). The nature of WLQs is still unclear so
far, though they are extensively studied \citep[see,][and the
references therein]{2011MNRAS.417..681L}. The optical-to-X-ray
spectral index, $\alpha_{\rm ox}$, of typical RQ AGNs ranges from
1.2 to 1.8, and the slope steepens with increasing UV luminosity
\citep{2005AJ....130..387S}. During the high state of PG 1407+265,
the optical-to-X-ray spectrum is flat ($\alpha_{\rm ox}$=1.09), and
there are larger fluctuations in the X-ray bands than the
UV bands \citep[i.e., the slope flattens with increasing
luminosity,][]{2006MNRAS.365..960G}, which seems to be different
from that in typical RQ AGNs. It is found that the UV variability
correlates with that of the X-ray emission in the high state, and
\citet{2006MNRAS.365..960G} suggested that the soft X-rays may be
the jet's non-thermal emission in the high state. This implies that
both the emission in the optical and X-ray bands may come from the
jets. It is interesting to note that the source PG 1407+265 may
probably have a relativistic jet moving toward us with a Doppler
factor of $\delta\gtrsim10$ from its radio morphology and the
variations in radio wavebands \citep{2003ApJ...591L.103B}. To our
knowledge, PG 1407+265 is the only RQ AGN with relativistic jets
reported in the literature so far.
\citet{2011ScChG..54..183C} fitted the multi-waveband SED of this
source with the homogeneous sphere jet model, and they found that
the jet is required to be in very extreme conditions. Thus, they
suggested that the radiation of the accretion disk may not be
negligible in the optical/UV bands of the SED for this source.



In this paper, we adopt an accretion disk/corona$+$jet model to
explore the disk-jet connection in these two unusual AGNs. In
Section 2, we give a brief summary of the jet and the disk-corona
emission models used to fit the observed SEDs. Section 3 contains
the results. The discussion and conclusions are given in Sections 4
and 5. Throughout this paper, a cosmology with $H_{0}=70$ km
s$^{-1}$Mpc$^{-1}$, $\Omega_{\rm m}=0.3$ and $\Omega_{\Lambda}=0.7$
is adopted.

\section{The model}

The homogeneous sphere jet model was widely used to explain the
observed SEDs of blazars, with which the main features of the
blazars' spectra can be well reproduced
\citep*[e.g.,][]{1998ApJ...509..608T, 2001ApJ...554..725T,
2008MNRAS.385..283C}. For RQ AGNs, the accretion disk/corona model
was developed to explain the optical/UV and X-ray spectra
\citep*[e.g.,][]{1991ApJ...380L..51H,1993ApJ...413..507H}. In this
work, we adopt the homogeneous sphere jet$+$accretion disk/corona
model to fit the observed SEDs of these two AGNs, and their physical
properties can be derived. We briefly describe the jet, and the
accretion disk/corona models employed in this work as follows.

\subsection{Jet model}

In this paper, we use one zone synchrotron$+$inverse Comptonization
(IC) models to calculate the jet emission of III Zw 2 and PG
1407+265. The model was widely used in blazar SED modeling
\citep[e.g.,][and the references therein]{2010MNRAS.402..497G}. The
emission region is assumed to be a homogeneous sphere with radius
$R$ embedded in the magnetic field $B$ \citep[see,][for the
details]{1998ApJ...509..608T, 2001ApJ...554..725T,
2008MNRAS.385..283C, 2010MNRAS.402..497G}. A broken power law
electron energy distribution,
\begin{equation}
N(\gamma )=\left\{ \begin{array}{ll}
                    N_{0}\gamma ^{-p_1}  &  \mbox{ $\gamma_{\rm min}\leq \gamma \leq \gamma_{0}$} \\
            N_{0}\gamma _{\rm 0}^{p_2-p_1} \gamma ^{-p_2}  &  \mbox{ $\gamma _{\rm 0}<\gamma\leq\gamma_{\rm max}$,}
           \end{array}
       \right.
\label{Ngamma}
\end{equation}
is assumed in our calculations.  Such a broken power law
distribution could be the result of the balance between the particle
cooling and escape rates in the blob \citep[see,][for the detailed
discussion]{1962SvA.....6..317K, 1994ApJ...421..153S,
1996ApJ...463..555I, 1998A&A...333..452K, 1998MNRAS.301..451G}. The
parameters of this model include, the radius $R$ of the blob, the
magnetic field strength $B$, electron break energy $\gamma_{0}$, the
minimum and maximum energy $\gamma_{\rm min}$, $\gamma_{\rm max}$,
of the electrons, the normalization of the particle number density
$N_{0}$, and the indexes $p_{1,2}$ of the broken power law particle
distribution. The observed spectrum of the jet can be calculated
when
the Lorentz factor $\Gamma=1/\sqrt{1-\beta^{2}}$, the viewing angle
$\theta$ of the jet with respect with the line of sight, and the
spectrum of the external seed photons, are supplied. The frequency
and luminosity can be transformed from the jet frame to
observational frame as: $\nu=\delta\nu'/(1+z)$ and $\nu
L_{\nu}=\delta^{4}\nu'L_{\nu'}'$, where the Doppler factor
$\delta=1/\left[\Gamma\left(1-\beta\cos\theta\right)\right]$, and
the prime represents the value measured in the jet frame. The
synchrotron self-absorption and the Klein-Nishina effect in the
inverse Compton scattering are properly considered in our
calculations
\citep[see,][]{1979rpa..book.....R,1970RvMP...42..237B}. Both the
self-synchrotron Compton (SSC) scattering and external Compton (EC)
scattering are included in the calculation of the Compton scattering
in the blob.

In this homogeneous sphere model, the jet power can be
calculated if all the physical quantities of the sphere are
specified,
\begin{equation}
L_{\rm jet}\simeq\pi R^{2}\beta \Gamma^{2}cU_{\rm tot}',
\label{l_jet}
\end{equation}
where the total energy density measured in the rest frame of the
blob,
\begin{displaymath}
U_{\rm tot}^\prime=U_{\rm e}^\prime+U_{\rm B}^\prime+U_{\rm
p}^\prime.
\end{displaymath}
The energy density for electrons $U_{\rm e}^\prime=m_{\rm
e}c^{2}\int N(\gamma)\gamma d\gamma$, while the proton energy
density $U_{\rm p}^\prime=U_{\rm e}^\prime(m_{\rm p}/m_{\rm
e})/\langle\gamma\rangle$ if charge neutrality for pure hydrogen
plasma is assumed \citep{1993MNRAS.264..228C, 2008MNRAS.385..283C}.
The broad waveband SED from high frequency radio emission to
$\gamma$-ray bands can be modeled with this homogeneous sphere model
\citep*[e.g.,][]{2008MNRAS.385..283C,2010MNRAS.402..497G}. The
emission from blazars in low frequency radio band may dominantly be
radiated from the inner conical jet near the black hole
\citep*[e.g.,][]{1979ApJ...232...34B,1981ApJ...243..700K,1985A&A...146..204G,1998ApJ...494..139J},
which is beyond the scope of this work. In this work, we use the
conventional homogeneous sphere model to fit the observed SEDs from
high frequency radio emission to $\gamma$-ray bands, which is
similar to most of the previous works
\citep*[e.g.,][]{2008MNRAS.385..283C,2010MNRAS.402..497G}.

\subsection{Accretion disk/corona model}

The observed UV/optical continuum emission of AGNs is thought to be
the thermal emission from the standard geometrically thin, optically
thick accretion disks
\citep*[e.g.,][]{1978Natur.272..706S,1982ApJ...254...22M,1989ApJ...346...68S},
while the observed power-law hard X-ray spectra of RQ AGNs are most
likely due to the inverse-Compton scattering of soft photons on a
population of hot electrons in the corona above the disk
\citep{1979ApJ...229..318G,1991ApJ...380L..51H,1993ApJ...413..507H}.
In the accretion disk-corona model, such soft photons are from the
cold disk, a fraction of which are Compton scattered by the hot
electrons in the corona above the cold disk to the hard X-ray energy
band. The disk-corona model was extensively explored in many
previous works
\citep*[e.g.,][]{1991ApJ...380L..51H,1993ApJ...413..507H,1994ApJ...436..599S,
2001ApJ...546..966K,2002ApJ...572L.173L,2009MNRAS.394..207C}. In
this disk-corona scenario, most of the gravitational energy of the
accreting matter is released in the cold disk, and a
fraction of which is transported into the corona probably by
magnetic fields. The magnetic fields generated in the cold disk are
strongly buoyant, and a substantial fraction of magnetic energy is
transported vertically to heat the corona above the disk with the
reconnection of the fields
\citep*[e.g.,][]{1998MNRAS.299L..15D,1999MNRAS.304..809D,2001MNRAS.328..958M,
2002MNRAS.332..165M,2009MNRAS.394..207C}. In this work, we adopt the
model given in \citet{2009MNRAS.394..207C} to calculate the spectrum
of the accretion disk/corona system. We summarize the accretion
disk-corona model in this sub-section.

The gravitational power dissipated in unit surface area of the
accretion disk is given by
\begin{equation}
Q_{\rm dissi}^{+}=\frac{3}{8\pi}\dot{M}\Omega_{\rm
k}(R)^{2}\left[1-\left(\frac{R_{\rm in}}{R}\right)^{1/2}\right],
\label{dissi}
\end{equation}
where $\dot{M}$ is the mass accretion rate of the disk, $\Omega_{\rm
k}(R)$ is the Keplerian velocity at radius $R$, and $R_{\rm
in}=3R_{\rm S}$ \citep{1973A&A....24..337S}. The Schwarzschild
radius $R_{\rm S}=2GM_{\bullet}/c^2$, where $M_{\bullet}$ is the
black hole mass. The accretion disk luminosity is
\begin{equation}
L_{\rm disk}=4\pi\int Q_{\rm dissi}^{+}RdR={\frac
{GM_{\bullet}\dot{M}}{2R_{\rm in}}}. \label{l_disk}
\end{equation}
In the absence of the corona, the surface temperature $T_{\rm s}$ of
the disk as a function of radius $R$ can be calculated with $\sigma
T_{\rm s}^4=Q_{\rm dissi}^{+}$, and the spectrum of the disk is
available by integrating blackbody emissivity $B_\nu(R)$ over radius
$R$ \citep{1973A&A....24..337S}. In this case, the spectrum of the
accretion disk can be derived when the black hole mass and the
accretion rate are specified.

In the accretion disk/corona system, the corona is assumed to be
heated by the reconnection of the magnetic fields generated by the
buoyancy instability in the disk. The power dissipated in the corona
is estimated with,
\begin{equation}
Q_{\rm cor}^{+}=p_{\rm m}v_{\rm p}={\frac {B^2}{8\pi}}v_{\rm p},
\label{Qcor}
\end{equation}
where $p_{\rm m}$ is the magnetic pressure in the disk, and $v_{\rm
p}$ is the velocity of the magnetic flux transported vertically in
the disk \citep{1998MNRAS.299L..15D}. The rising speed $v_{\rm p}$
is assumed to be proportional to the internal Alfv\'{e}n velocity,
i.e., $v_{\rm p}=bv_{\rm A}$, in which $b$ is of the order of unity
for extremely evacuated magnetic tubes. We adopt $b=1$ in all our
calculations of this work.

The soft photons from the disk are Compton scattered by the hot
electrons in the corona to X-ray bands, and about half of the
scattered photons are intercepted by the disk. The reflection albedo
$a$ is relatively low, $a\sim0.1-0.2$, and most of the incident
photons from the corona are re-radiated as blackbody radiation
\citep[e.g.,][]{1999MNRAS.303L..11Z}. Thus, the energy equation for
the cold disk is
\begin{equation}
Q_{\rm dissi}^{+}-Q_{\rm cor}^{+}+\frac{1}{2}(1-a)Q_{\rm
cor}^{+}=\frac{4\sigma T^{4}_{\rm disk}}{3\tau}, \label{eneq}
\end{equation}
where $T_{\rm disk}$ is the effective temperature in the mid-plane
of the disk, and $\tau=\tau_{\rm es}+\tau_{\rm ff}$ is the optical
depth in vertical direction of the disk. In this work, we adopt
$a=0.15$ in all our calculations.

The detailed physics for generating magnetic fields in the accretion
disk is still quite unclear, and there are three different magnetic
stress tensors were usually adopted,
\begin{equation}
 \tau_{\rm r\varphi}=p_{\rm m}=\left\{ \begin{array}{ll}
        \alpha p_{\rm tot} \\
         \alpha p_{\rm gas} \\
         \alpha \sqrt{p_{\rm gas}p_{\rm tot}}, \\
\end{array} \right.
\label{viscosity}
\end{equation}
where $\alpha$ is the viscosity parameter
\citep[see][]{1973A&A....24..337S, 1981ApJ...247...19S,
1984ApJ...287..761T}.
The parameters of the disk-corona model include the black hole mass
$M_{\bullet}$, the dimensionless mass accretion rate $\dot{m}$
($\dot{m}=\dot{M}/\dot{M}_{\rm Edd}$, $\dot{M}_{\rm Edd}\equiv
L_{\rm Edd}/\eta_{\rm rad}c^2$, and the conventional radiative
efficiency $\eta_{\rm rad}=0.1$ is adopted), and the viscosity
parameter $\alpha$ \citep*[see][for the
details]{2009MNRAS.394..207C}.


\section{Results}

We search the literature for the observational data of the
multi-waveband SEDs of these two sources from radio to X-ray bands
\citep[][and
\emph{NED}\footnote{http://ned.ipac.caltech.edu/}]{2002MNRAS.335..177S,1988A&A...198...16K}.
Obviously, simultaneous broad band SEDs are desired to be used to
model the disk-jet systems in these sources. For III Zw 2, we search
the literature and collect a broad band SED, which is
quasi-simultaneous within half a year (see Table \ref{iii_ta_sed}).
This source shows long-term variability on time scale of a few
years, which is longer than the temporal span of the
quasi-simultaneous data. Therefore, we mainly adopt this SED for our
model fitting. For the source PG 1407+265, we collect and list all
X-ray observations in Table \ref{pg_ta_x-ray}. We note that the
GINGA observation (Jun. 1987) is roughly simultaneous with IR and
optical observations (see Table \ref{pg_ta_sed} and
\ref{pg_ta_x-ray}). Unfortunately, only 2-10 keV flux is available
without spectral index. It can be seen that the 2-10 keV flux of
observation on Jan. 17, 1981 by EINSTEIN is similar to that of GINGA
observation. We therefore use EINSTEIN observation data in the
quasi-simultaneous spectrum of this source (see Table
\ref{pg_ta_x-ray}).

\subsection{III Zw 2}

\begin{figure}
\epsscale{0.8} \plotone{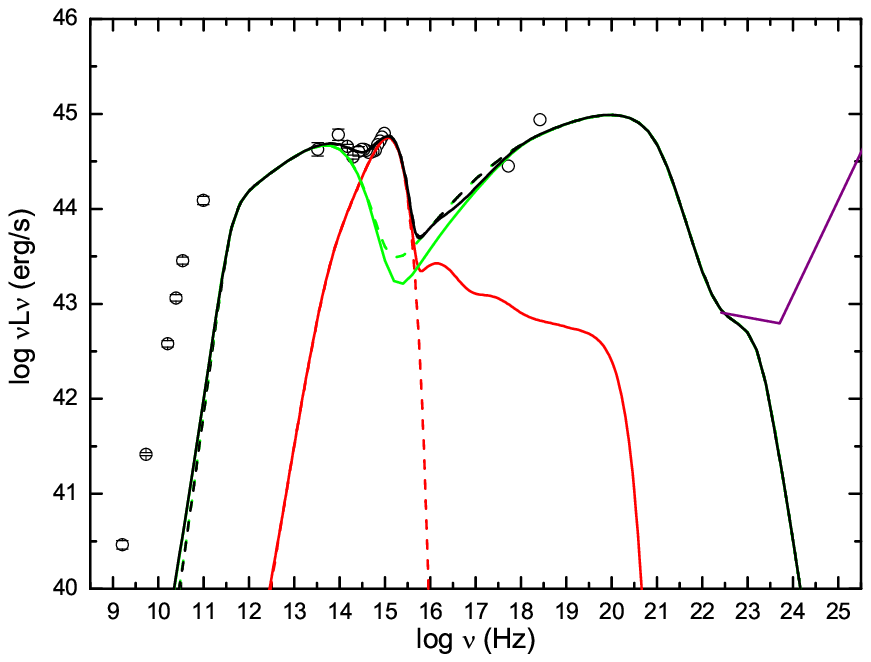} \caption{The quasi-simultaneous SED of III Zw 2 (see Table \ref{iii_ta_sed} for references of the data). The method of least squares is used to fit SED. The green solid line represents
the emission from the jet, while the red solid line is for the
disk/corona emission. The red dashed line represents the spectra
from the bare accretion disk, while the green dashed line is for the emission from the jet in this case. The black line is the sum of the emission from the jet and the accretion disk/corona. The purple
line represents the sensitivity of $\gamma$-ray detector \emph{Fermi}/LAT
\citep{2004ASSL..304..361M}. The parameters model calculations see Table \ref{iii_ta}.}
\label{iii_sim}
\end{figure}

We search the literature and collect the quasi-simultaneous
SED of III Zw 2 from radio to X-ray bands, which is plotted in
Figure \ref{iii_sim} (the data are listed in Table
\ref{iii_ta_sed}).
\citet{2002MNRAS.335..177S} estimated the central black hole
mass $M_{\bullet}=10^{9}M_{\odot}$ from the width of the broad-line
H$\beta$ of this source. Superluminal motion of the jet component
with apparent velocity $\beta_{\rm app}=1.25\pm0.09$ in III Zw 2 has
been detected with VLBA observations \citep{2000A&A...357L..45B}.
\citet{2003ApJ...599..185P} derived the inclination angle of the
emission line disk $\theta=12^{\circ}\pm5^{\circ}$ for III Zw 2 from
its broad emission line profile. Assuming the jet to be
perpendicular to the emission line disk, the Lorentz factor of the
jet, $\Gamma\simeq2.06$, and the Doppler factor, $\delta\simeq3.35$,
are derived.

In the calculations of the spectra from the jets, we have considered
both the SSC and EC mechanisms. The external soft seed photons from
the broad line region (BLR) are considered in our calculations. The
VLBA observations and the variability in radio bands imply that the
location of the blob is within the BLR
\citep[see][]{2000A&A...357L..45B}. The radius of the emission
region is estimated with the minimum variability time scale
$R\approx ct_{\rm var}\delta/(1+z)\approx1.5\times10^{16}$ cm
\citep{1997AJ....114..565J}. \citet{1988A&A...198...16K} estimated
the size of the BLR $\approx10^{18}$ cm, and the photon energy
density within the BLR $U_{\rm ext}\approx3.8\times10^{-4}$ erg
cm$^{-3}$. Thus, the multi-waveband spectrum of the jet can be
calculated if the values of the parameters, $N_{0}$, $B$,
$\gamma_{0}$, $\gamma_{\rm min}$, $p_{1}$ and $p_{2}$, are provided.
Besides the spectrum of the jet, we need to calculate the spectra
from the accretion disk/corona system in this source. First, we
consider the simplest case, i.e., a bare accretion disk without a
corona. In this case, the spectrum of the accretion disk is only
dependent on the mass accretion rate (see Section 2).
From Figure \ref{iii_sim}, it
can be seen that the quasi-simultaneous SED seems to show two bumps
at IR and UV bands respectively. The UV
bump is a typical character of RQ AGN, which can be explained as the
thermal emission from an accretion disk emission. We model the UV
bump as disk emission, and jet emission accounts for the IR bump. In
the model fitting the quasi-simultaneous SED, the least squares
method is adopted
to derive the best fitted model parameters. The model parameters,
the accretion rate $\dot{m}$, the minimum electron energy
$\gamma_{\rm min}$, and the magnetic field strength $B$, are
carefully tuned in order to obtain the best model fitting to the
quasi-simultaneous SED. The best fitting parameters are listed in
Table \ref{iii_ta} (see Figure \ref{iii_sim}). With the values of
the parameters determined from the fitting of the SEDs, the jet
power can be calculated with Equation (\ref{l_jet}). In this case,
jet power is $L_{\rm jet}^{\rm sim}\approx1.5\times10^{45}$ erg
s$^{-1}$ and the accretion disk luminosity $L^{\rm sim}_{\rm
disk}\approx1.1\times10^{45}$ erg s$^{-1}$.

We also repeat our calculations by adopting the accretion
disk/corona model instead of the bare accretion disk model as
described above. There are different choices of magnetic stresses
(see Equation \ref{viscosity}). Our results show that the
contribution to the X-ray continuum spectra in this source is always
negligible (see Table \ref{iii_ta}).

We plot the sensitivity of the detector of \emph{Fermi}/LAT in the
Figure \ref{iii_sim} \citep[purple lines,][]{2004ASSL..304..361M}.
It can be seen that if the optical/UV emission is dominantly
from the accretion disk, the predicted $\gamma$-ray flux is below
the sensitivity of \emph{Fermi}/LAT.

\subsection{PG 1407+265}

The quasi-simultaneous SED of PG 1407+265 shows clear evidence of an
optical-UV bump (see Figure \ref{pg}). \citet{2010MNRAS.404.2028H}
estimated the black hole mass of this source to be
$6\times10^{9}M_{\odot}$. There are no (quasi-)simultaneous
optical/IR spectra corresponding to the high state in the X-ray
band. Thus, we have to limit our model fittings to the
quasi-simultaneous SED in the low state. \citet{2003ApJ...591L.103B}
suggested that there is a relativistic jet in PG 1407+265, and the
Doppler factor of the jet is estimated as $\delta\gtrsim10$. We
adopt $\delta=10$ in our calculations for this source. The procedure
of fitting its SED is the same as that for III Zw 2. We find that
the optical/UV component in the SED may probably be from the
accretion disk, while the origin of the IR component is still
uncertain. There are two possibilities: the synchrotron emission
from the jet, or the thermal radiation from the dust torus
irradiated by the radiation of the central engine. In the case of
the synchrotron emission, the minimum electron energy can be
constrained, $\gamma_{\rm min}\approx13.5$, if equipartition between
magnetic field and electron$+$proton energy is assumed. The jet
power is $L_{\rm jet}\approx1.1\times10^{46}$ erg s$^{-1}$, while
the disk luminosity is $L_{\rm disk}\approx6.8\times10^{46}$ erg
s$^{-1}$. If the IR component is from the dust torus, our
calculations require the minimum electron energy $\gamma_{\rm
min}\approx4.6$, and the jet power is $L_{\rm
jet}\approx3.3\times10^{44}$ erg s$^{-1}$, which is significantly
lower than the disk luminosity. All the model parameters adopted for
different cases are listed in Table \ref{pg_ta}.

In order to explore the origin of the X-ray emission in PG 1407+265,
we also model the SED of this source by including the contribution
of the corona above the disk. In our calculations, we adopt
different magnetic stresses respectively, and find that the
contribution of the corona to the X-ray spectrum is always
negligible, unless the stress $\tau_{\rm r\varphi}=p_{\rm m}=\alpha
p_{\rm tot}$ is adopted with an relatively high viscosity parameter:
$\alpha\sim 0.5$ (see Figure \ref{pg_ma}).

\subsection{Blazar sequence}

\begin{figure}
\epsscale{1} \plotone{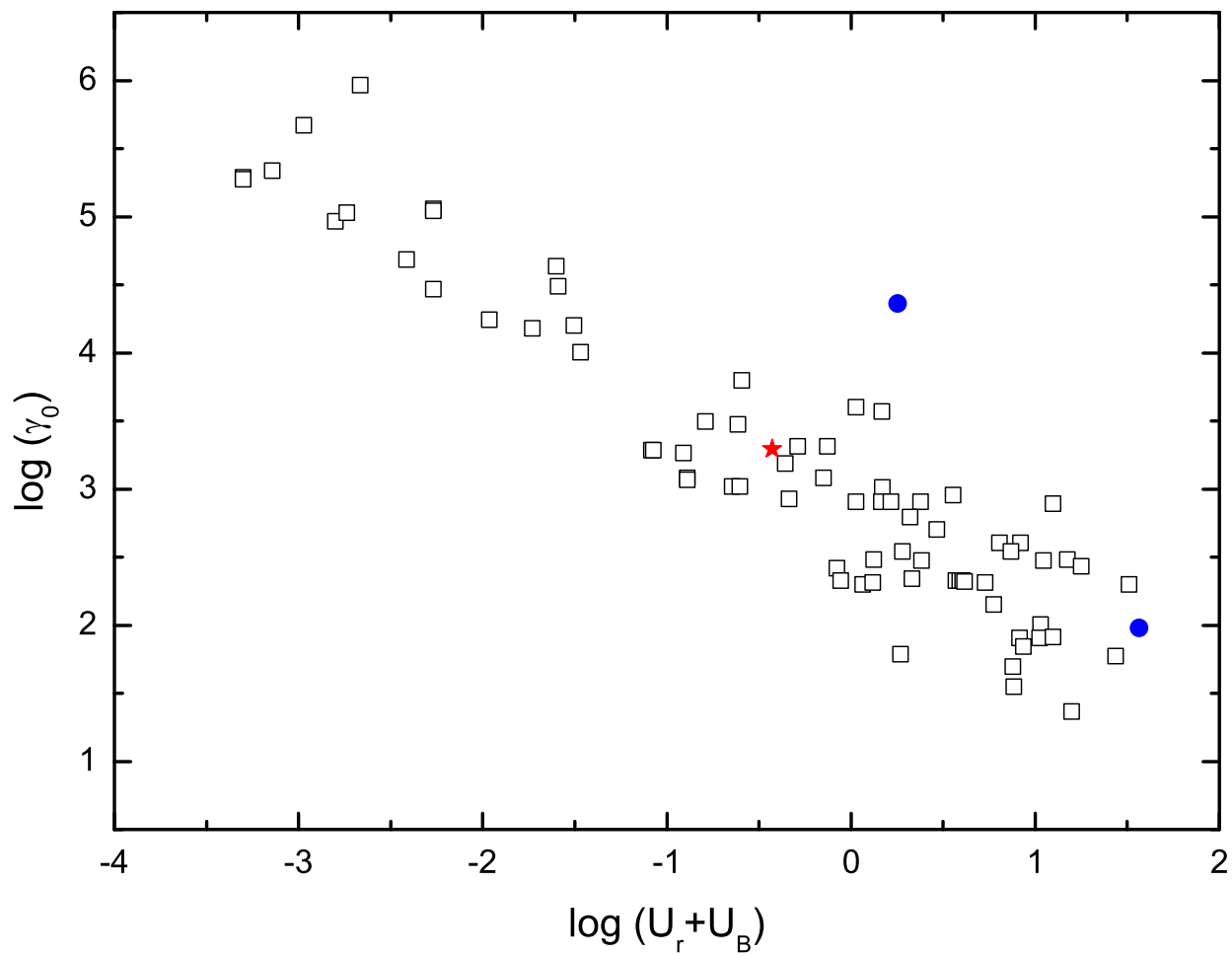} \caption{The blazar sequence:
$\gamma_{\rm 0}$ vs. U$_{\rm tot}'$. The opened squares represent the
blazars taken from \citet{2008MNRAS.385..283C}. The red stars
represent III Zw 2, while
the blue dots indicate PG 1407+265. The dot with a higher
$\gamma_{0}$ corresponds to the case with the X-ray emission
dominated by the synchrotron emission from the jet, while the lower
one corresponds to the case of the X-ray emission emission dominated
by the inverse Compton scattering in the jet.} \label{sequence}
\end{figure}

\begin{figure}
\epsscale{1} \plotone{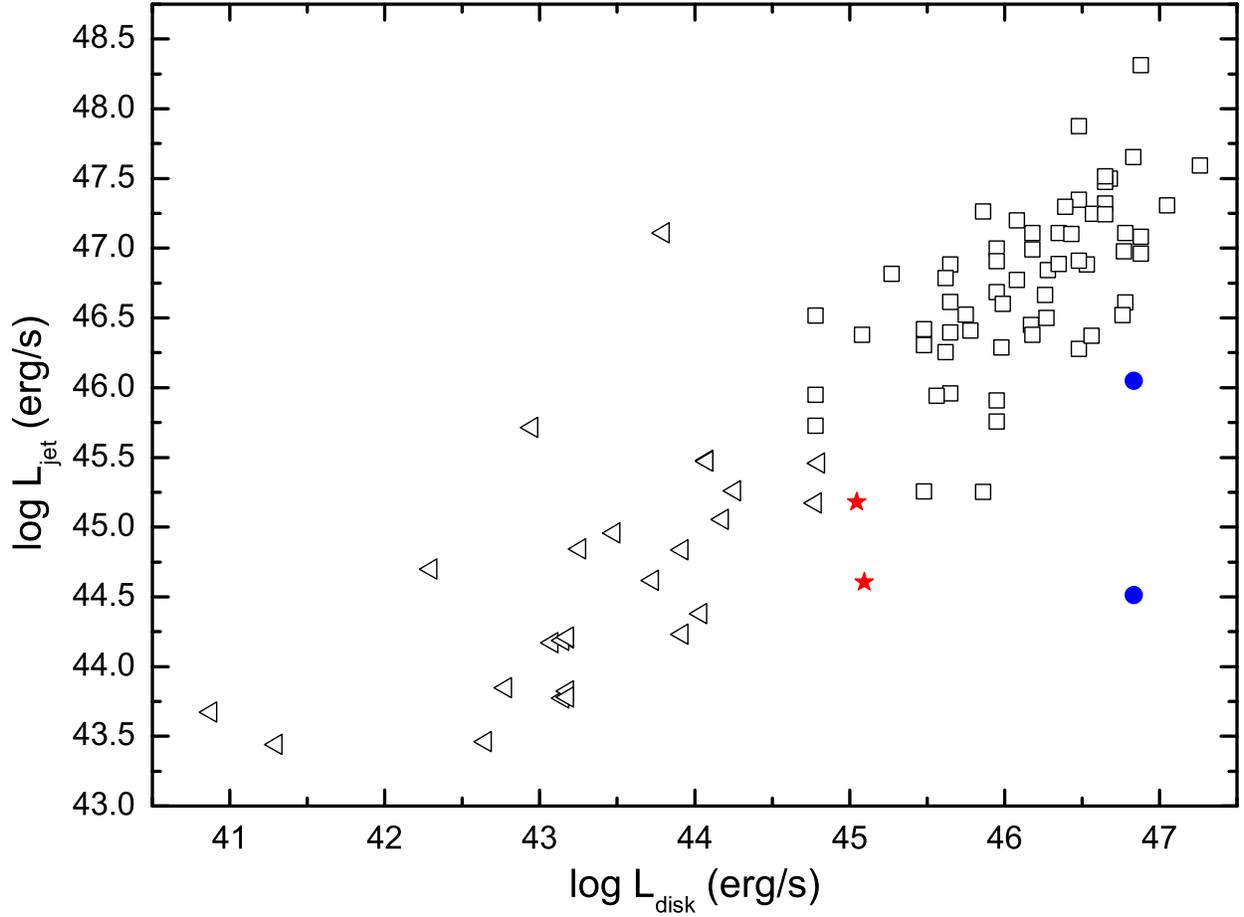} \caption{Jet power $L_{\rm jet}$
vs. disk luminosity $L_{\rm disk}$. The black triangles/squares
present the blazars, while the triangles are for the sources only
with upper limits on the disk luminosity
\citep[see,][]{2010MNRAS.402..497G}. The red stars represent III Zw
2, while the blue dots
are for PG 1407+265. The dot with lower jet power corresponds to
the case of the X-ray emission dominated by the synchrotron emission
from the jet, while the higher one is for case of the X-ray emission
dominated by the inverse Compton scattering in the jet.}
\label{ld_ljet}
\end{figure}

We compare these two sources with the well studied blazar sample
\citep{2008MNRAS.385..283C}. In Figure \ref{sequence}, the blazar
sequence is plotted, and we find that these two sources roughly
follow the correlation defined by the blazar sample, except the
source PG 1407+265 if the IR component is from the dust torus and
the X-ray emission is the synchrotron emission from the jet. We also
compare the relation of jet power and disk luminosity with that of
the blazar sample in Figure \ref{ld_ljet}, in which the blazar
sample is taken from \citet{2010MNRAS.402..497G}.




\section{Discussion}

The simultaneous SEDs either in the high or low states are
not available for III Zw 2, and only a quasi-simultaneous SED is
obtained by searching the data in literature.
This quasi-simultaneous SED
shows a big blue bump, which is a typical character of RQ quasar and
can be modeled quite well by the accretion disk model (see Figure
\ref{iii_sim}). Compared with RQ AGNs, it is known that
relatively high polarization in the optical waveband is observed in
blazars, because the emission in this waveband is dominated by the
synchrotron emission from the relativistic jets. We note that almost
no polarization has been observed in the optical waveband for III Zw
2 \citep[$\sim0.28\pm0.19\%$,][]{1990ApJS...74..869B}, and the
amplitude of its optical variability is smaller than the radio and
X-ray emission \citep{2002MNRAS.335..177S}, which imply that the
optical/UV emission in this source may not be dominantly from the
jet.
We estimate the thermal timescale of the accretion disk
\citep[assuming black hole spin $a=0$, and using the same
method as that in ][]{2008ApJ...677..884L}, $\tau_{\rm
thermal}\approx2-5$ yr for this source in the low or the high
states. This is roughly consistent with the observed optical/UV
variability timescales \citep{2002MNRAS.335..177S}.
\citet{2010RAA....10..707C} showed that the predicted $\gamma$-ray
emission can be be detected by \emph{Fermi}/LAT, if the jet emission
is responsible for the optical/UV continuum emission. However, the
model fittings to the simultaneous SED of this source in this work
show that the UV/optical emission originates from the accretion disk
(see Figure \ref{iii_sim}), and the $\gamma$-ray emission from the
jet is below the sensitivity of \emph{Fermi}/LAT. We note that III
Zw 2 is not included in either 1LAC \citep[the first LAT AGN
Catalog,][]{2010ApJ...715..429A} or 2LAC \citep[the second LAT AGN
Catalog,][]{2011ApJ...743..171A}. We suggest that the optical/UV
continuum spectra of this source may probably be dominantly emitted
from the accretion disk. This issue can be sorted out if the
simultaneous multi-waveband observations on this source both in low
and high states are performed in the future.

We also calculate the X-ray spectrum of the corona above the disk,
and find that the contribution of the corona to the X-ray emission
is always negligible, i.e., the X-ray continuum spectra are mainly
emitted from the relativistic jet in this source. \citet{2007ApJ...658..815S} investigated the radio loudness of a
sample with 199 sources consisting of broad-line radio galaxies
(BLRGs), radio-loud quasars (RLQs), Seyfert galaxies (SGs),
low-ionization nuclear emission-line region galaxies (LINERs), and
Fanaroff-Riley type I radio galaxies (FR I RGs).
They found that there are two distinct, approximately parallel
tracks in the plot of $\log~R-\log~\lambda$ (the Eddington ratio
$\lambda\equiv L_{\rm bol}/L_{\rm Edd}$, where $L_{\rm bol}$ is the
bolometric luminosity and $L_{\rm Edd}$ is the Eddington
luminosity). We compare
III Zw 2 with the results in \citet{2007ApJ...658..815S}, and find
that this source locates between the RL and RQ tracks (Eddington
radio: $\lambda^{\rm low}\approx0.019$, $\lambda^{\rm
high}\approx0.072$, $\lambda^{\rm sim}\approx0.039$; radio-loudness
$R\sim100-200$, \citet{1996ApJ...471..106F,2000A&A...357L..45B}).

Strong Fe K$\alpha$ line emission with equivalent width
EW$\simeq220$ {\AA} has been detected in III Zw 2
\citep{2005A&A...435..449J}. It roughly follows the relation between
the equivalent width and luminosity for RQ quasars, of which the
X-ray continuum emission is supposed to be from the accretion
disk/corona system \citep{1997ApJ...488L..91N}. Our results suggest
that only a small fraction of the observed X-ray continuum is from
the corona of the accretion disk, which implies that this source in
fact deviates significantly from the relation between the equivalent
width and luminosity defined by the RQ quasar sample if the X-ray
continuum emission from the jet is properly subtracted in this
source. The X-ray emission from the corona is obviously not enough
to power such a strong Fe K$\alpha$ in this source, and therefore
the detailed studies on X-ray reflection geometry in III Zw 2 may be
necessary for resolving this issue, which is beyond the scope of
this work.

\begin{figure}
\epsscale{0.8} \plotone{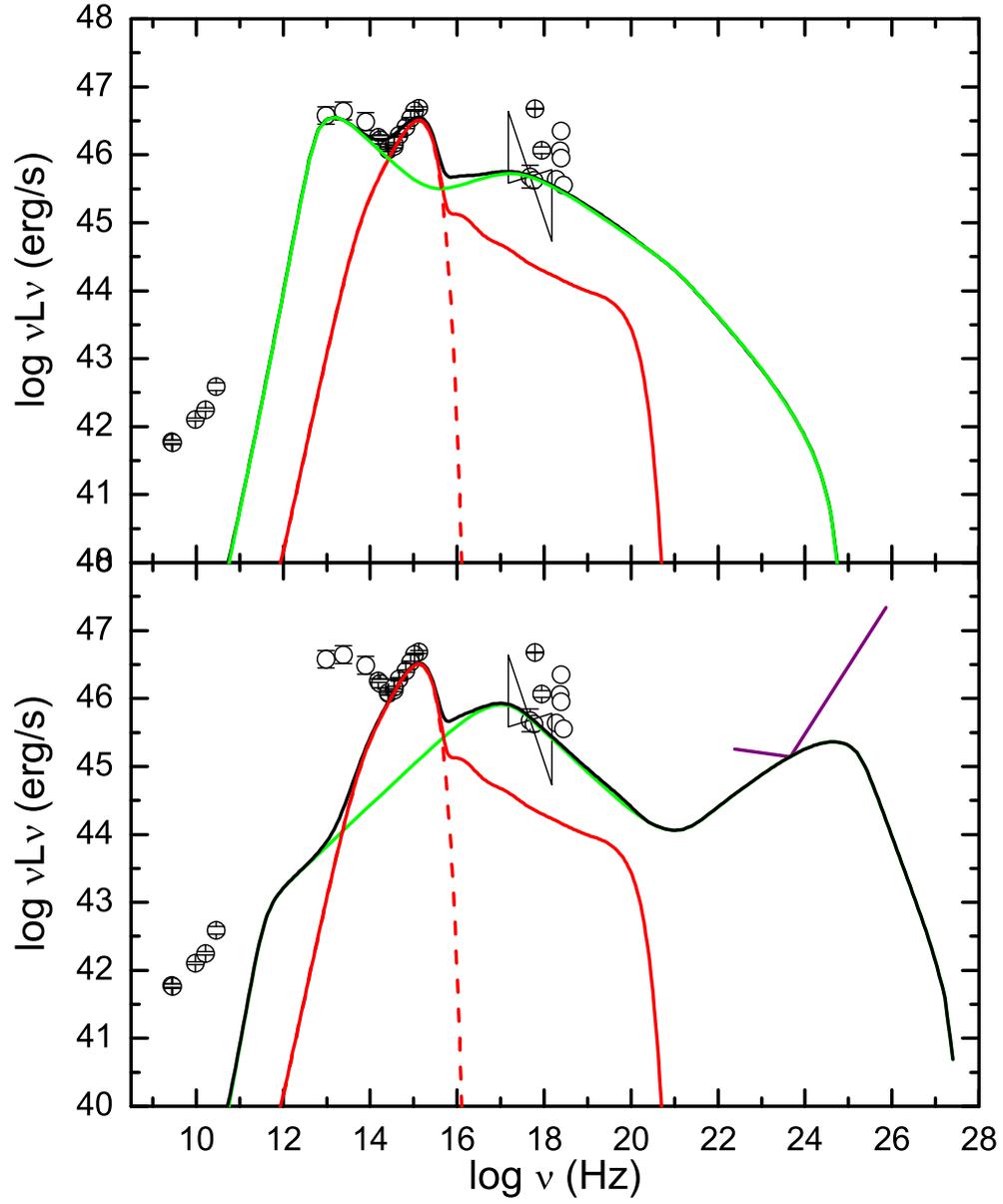} \caption{The same as Figure
\ref{iii_sim}, but for the source of PG 1407+265. The method of least squares is used in fitting.} \label{pg}
\end{figure}

The SED of PG 1407+265 exhibits a clear component in optical/UV
bands, which can be well fitted by the accretion disk model (see,
Figure \ref{pg}). \citet{2011ScChG..54..183C} fitted the multi-band
SED of this source using the one-zone jet model, and found that
extreme values of some model parameters are required.
Similar to III Zw 2, almost no polarization in optical
emission has been detected in this source
\citep[$\sim0.24\pm0.16\%$,][]{1990ApJS...74..869B}, which may also
suggest that most of the optical-UV emission may not come from the
jet.

The origin of the X-ray emission in this source is somewhat
uncertain, which could be dominantly either from the jet or the
corona above the disk. We firstly adopt the jet$+$accretion
disk/corona model to fit its SEDs. In this case, our results show
that the X-ray emission is dominantly from the jet. Unlike III Zw 2,
a component in IR wave bands has been observed in PG 1407+265, which
may originate from the synchrotron emission from the jet, or the
thermal radiation from the dust torus irradiated by the radiation of
the central engine. In the case of the synchrotron emission,
the derived jet power is $L_{\rm jet}\approx1.1\times10^{46}$ erg
s$^{-1}$, which is lower than the disk luminosity ($L_{\rm
disk}\approx6.8\times10^{46}$ erg s$^{-1}$). If the IR component is
alternatively assumed to be emitted from the dust torus, our
calculations indicate that the jet power $L_{\rm
jet}\approx3.3\times10^{44}$ erg s$^{-1}$, which is also
significantly lower than the disk luminosity.
\citet{2003ApJ...591L.103B} estimated the lower limit on the jet
power in this source, $\sim 7\times 10^{43}$~erg s$^{-1}$, by
assuming minimum of the total energy density, in which the minimum
magnetic field strength $B_{\rm min}$ is adopted. Considering that
the realistic field strength could be significantly higher than
$B_{\rm min}$, their estimate of the lower limit on jet power is
consistent with our results in the case of the
IR emission being from the torus. The opening angle of the dust
torus can be estimated from the ratio of IR to optical/UV fluxes by
assuming that the radiation of the disk is absorbed by the dust
torus and re-radiates in IR wavebands \citep[see][for the detailed
discussion]{2005ApJ...619...86C}. The ratio of IR to optical/UV
emission $\sim0.6$ for PG 1407+265, and the torus opening angle
$\sim50^{\circ}$ is inferred, which is typical for RQ AGNs
\citep{1994ApJS...95....1E,2005ApJ...619...86C,2011ApJS..196....2S}.
We find that the X-ray emission is dominated by the synchrotron
emissions from the jet (see the lower panel of Figure \ref{pg}), and
this source should be a ``high peak frequency blazar". This source
is luminous, and it is quite different compared with normal blazars,
because the blazar sequence expects a high peak frequency blazar
when its luminosity is low
\citep{1998MNRAS.299..433F,1998MNRAS.301..451G,2011ApJ...735..108C}.

\begin{figure}
\epsscale{1} \plotone{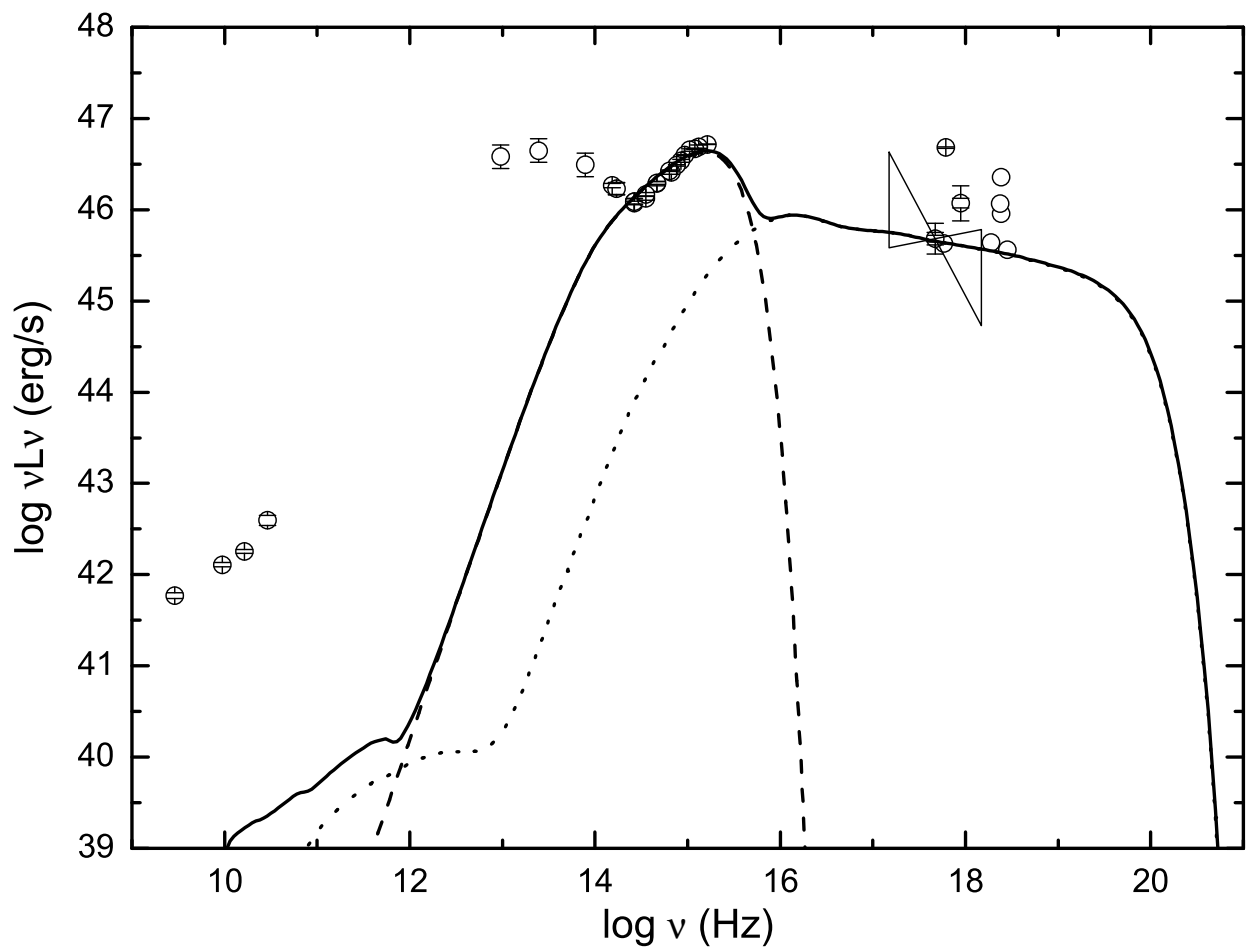} \caption{The same as Figure
\ref{pg}, but the SED is fitted with the accretion disk/corona
model. The magnetic stress $\tau_{\rm r\varphi}=\alpha p_{\rm tot}$
is used, and the best-fitted model parameters are: $\dot{m}=0.3$ and
$\alpha=0.5$. The solid line represents the spectra from the
accretion disk/corona system. The dashed line represents the disk
spectra, while the dotted line is for the spectra from the
corona.} \label{pg_ma}
\end{figure}

In order to investigate the origin of the X-ray emission in this
source, we alternatively adopt the accretion disk/corona model to
fit the SED for this source. We find that the spectra in the
optical/UV and X-ray bands can be fairly well fitted with the
magnetic stress $\tau_{\rm r\varphi}=\alpha p_{\rm tot}$, provided
$\dot{m}=0.3$ and $\alpha=0.5$ are adopted (see Figure \ref{pg_ma}).
However, we note that the variability of the X-ray emission
in this source can be around an order of magnitude between its low
and high states (see Figure \ref{pg_ma}), which is more violent than
normal RQ quasars. Future simultaneous multi-band
(UV/optical$+$X-ray) observations on this source will help resolve
this issue.
In the accretion disk/corona model, the variability in the X-ray
band may lead to variable optical-UV emission, and vice versa. The
observational data show strong variability in the X-ray band of this
source, while no evidence of similar variability is found in the
optical/UV band. Further simultaneous multi-waveband observations is
expected to attack the nature of the X-ray emission in this source.
We compare PG 1407+265 with the results in
\citet{2007ApJ...658..815S}, and find that this source locates in
the RQ track (Eddington radio $\lambda\approx0.33$; radio-loudness
$R\approx0.44,3.43$,
\citet{1987ApJ...323..243W,1989AJ.....98.1195K}). Therefore, PG
1407+265 could be a typical RQ quasar, but contains relativistic
weak jets.

The nature of WLQs is still unclear. \citet{2011MNRAS.417..681L}
proposed that the accretion disk temperature decreases with
increasing black hole mass, and the fraction of the photons with the
energy that can ionize the BLR decreases. This may account for the
weak line emission in some AGNs. They suggested that, the quasar
will be lineless if the fraction of the accretion disk emission
above the frequency $\nu=3.29\times10^{15}$ Hz is less than $0.01$
($L_{\nu>3.29\times10^{15}\rm Hz}/L\lesssim0.01$), while a weak line
quasar appears if  $L_{\nu>3.29\times10^{15}\rm Hz}/L\lesssim0.1$.
PG 1407+265 is the first WLQ studied in detail
\citep{1995ApJ...450..585M}. We calculate the fraction of
disk-corona emission above the frequency,
$L_{\nu>3.29\times10^{15}\rm Hz}/L\approx0.08-0.1$, which satisfies
the WLQ criterion suggested by \citet{2011MNRAS.417..681L}.

We compare these two sources with the well studied blazar sample
\citet{2008MNRAS.385..283C}. We find that these two sources roughly
follow the blazar sequence defined by the blazar sample, and also
compare the relation of jet power and disk luminosity with that of
blazars. We find that these two sources do not deviate much from the
blazars, except the source PG 1407+265, if its IR component
is assumed to be from the dust torus and the X-ray emission is the
synchrotron emission from the jet.


\section{Summary}

We summarize the main conclusions of this paper as follows.

1. The optical/UV emission of III Zw 2 is dominantly from the
accretion disk.
The X-ray emission should be
dominated by the radiation from the jet, while the contribution of
the disk corona is negligible.

2. The predicted $\gamma$-ray emission is below the sensitivity of \emph{Fermi}/LAT, provided the UV is emission originated from the accretion disk.


3. The optical/UV component in the SED of PG 1407+265 can be well
modeled as the emission from the accretion disk, while the IR
component is attributed to the thermal radiation from the dust torus
with an opening angle $\sim 50^\circ$. If the X-ray continuum
emission is dominantly from the jet, the source should be a ``high
peak frequency blazar", which obviously deviates the normal blazar
sequence.

4. The observed SED can also be fitted quite well by the accretion
disk/corona model with the viscosity parameter $\alpha=0.5$. The
spectrum of the accretion disk/corona in the WLQ PG 1407+265
satisfies the WLQ criterion suggested by
\citet{2011MNRAS.417..681L}.

\acknowledgments

We thank the anonymous referee for insightful comments and
constructive suggestions. This work is supported by the NSFC (grants
10821302, 10833002, 11173043, 11133006, 10973034, 10903025, 11078008, and 11103060), the National Basic Research
Program of China (grant 2009CB824800), the Science and Technology
Commission of Shanghai Municipality (10XD1405000), and the CAS/SAFEA
International Partnership Program for Creative Research Teams
(KJCX2-YW-T23).

\begin{deluxetable}{crrrrrrrcccc}
\setlength{\tabcolsep}{0.07in}
\tablewidth{0pt}
\tabletypesize{\footnotesize}
\tablecaption{Parameters of III Zw 2
}
\tablehead{
\colhead{Model}&
\colhead{$B$(Gs)}&
\colhead{$N_{0}$}&
\colhead{$p_{1}$}&
\colhead{$p_{2}$}&
\colhead{$\gamma_{0}$}&
\colhead{$\gamma_{\rm min}$}&
\colhead{$\dot{m}$}&
\colhead{$\alpha$}&
\colhead{$\tau_{\rm r\varphi}$}&
\colhead{$L_{\rm jet}^{\dag}$}&
\colhead{$L_{\rm disk}^{\dag}$}
}
\startdata
1  & 2.08 & 7.02$\times10^{5}$ & 2.3 & 7.0 & 1963 & 20.0 & $5.1\times10^{-3}$ & ... & ...                                   & 1.51 & 1.12 \\
2  & 2.08 & 7.02$\times10^{5}$ & 2.3 & 7.0 & 1963 & 58.1 & $1.0\times10^{-2}$ & 0.2 & $\alpha\sqrt{p_{\rm gas}p_{\rm tot}}$ & 0.40 & 1.25
\enddata
\tablecomments{$^{\dag}$ $L_{\rm jet}$ and $L_{\rm disk}$ are in unit of 10$^{45}$ erg/s. $B$ is the magnetic field, $N_{0}$ the normalized
number density, $p_{1,2}$ the indexes of the broken power law
electron energy distribution, $\gamma_{0}$ the peak electron
energy, $\gamma_{\rm min}$ the minimum electron energies,
$\dot{m}$ the dimensionless accretion rate,
$\alpha$ the viscosity parameter.
Model 1 indicates the dashed line in Figure \ref{iii_sim}.
Model 2 is for the solid line in Figure \ref{iii_sim}.}\label{iii_ta}
\end{deluxetable}

\begin{deluxetable}{cccc}
\setlength{\tabcolsep}{0.07in}
\tablewidth{0pt}
\tabletypesize{\footnotesize}
\tablecaption{quasi-simultaneous SED data of III Zw 2}
\tablehead{
\colhead{$\log\nu$}&
\colhead{$\log f_{\nu}$(Jy)}&
\colhead{Observation Date}&
\colhead{Ref.}
}
\startdata
 9.166 & -1.000$\pm$0.043 & Jan. 1978 & S78\\
 9.689 & -0.569$\pm$0.016 & Jan. 1978 & S78\\
10.169 &  0.114$\pm$0.033 & Jan. 1978 & S78\\
10.352 &  0.415$\pm$0.033 & Jan. 1978 & S78\\
10.497 &  0.663$\pm$0.038 & Jan. 1978 & S78\\
10.954 &  0.839$\pm$0.044 & Jan. 1978 & S78\\
13.480 & -1.150$\pm$0.070 & Jul. 1977 & N79\\
13.934 & -1.450$\pm$0.060 & Jul. 1977 & N79\\
14.134 & -1.770$\pm$0.020 & Jul. 1977 & N79\\
14.260 & -2.010$\pm$0.020 & Jul. 1977 & N79\\
14.380 & -2.070$\pm$0.050 & Jul. 1977 & N79\\
14.450 & -2.120           & Jul. 1977 & N79\\
14.500 & -2.170           & Jul. 1977 & N79\\
14.550 & -2.230           & Jul. 1977 & N79\\
14.600 & -2.300           & Jul. 1977 & N79\\
14.650 & -2.350           & Jul. 1977 & N79\\
14.700 & -2.390           & Jul. 1977 & N79\\
14.750 & -2.430           & Jul. 1977 & N79\\
14.800 & -2.420           & Jul. 1977 & N79\\
14.850 & -2.430           & Jul. 1977 & N79\\
14.900 & -2.440           & Jul. 1977 & N79\\
14.950 & -2.450           & Jul. 1977 & N79\\
17.680 & -5.523           & Aug. 1977 & K88\\
18.379 & -5.732           & Aug. 1977 & K88
\enddata
\tablecomments{The quasi-simultaneous spectra of III Zw 2 as plotted in Figure \ref{iii_sim}. The reference S78 refers to \citet{1978ApJ...222L..91S}, N79: \citet{1979ApJ...230...79N}, K88: \citet{1988A&A...198...16K}.}\label{iii_ta_sed}
\end{deluxetable}

\begin{deluxetable}{cccc}
\setlength{\tabcolsep}{0.07in}
\tablewidth{0pt}
\tabletypesize{\footnotesize}
\tablecaption{SED data of PG 1407+265}
\tablehead{
\colhead{$\log\nu$}&
\colhead{$\log f_{\nu}$(Jy)}&
\colhead{Observation Date}&
\colhead{Ref.}
}
\startdata
 9.17 & -2.06$\pm$0.03 & Sep. 1993          & B96           \\
 9.69 & -2.24$\pm$0.02 & Sep. 1993          & B96           \\
 9.93 & -2.33$\pm$0.02 & Sep. 1993          & B96           \\
10.17 & -2.23$\pm$0.06 & Sep. 1993          & B96           \\
12.69 & -0.77$\pm$0.13 & Jun. 1996          & H03           \\
13.10 & -1.11$\pm$0.13 & Jun. 1996          & H03           \\
13.61 & -1.77$\pm$0.13 & Jun. 1996          & H03           \\
13.90 & -2.29$\pm$0.02 & Apr. 1988          & E94,\emph{NED}\\
13.95 & -2.37$\pm$0.06 & Apr. 1988          & E94,\emph{NED}\\
13.95 & -2.37$\pm$0.07 & Apr. 1988          & E94,\emph{NED}\\
14.13 & -2.70$\pm$0.03 & Apr. 1988          & E94,\emph{NED}\\
14.13 & -2.72$\pm$0.02 & Apr. 1988          & E94,\emph{NED}\\
14.13 & -2.70$\pm$0.03 & Apr. 1988          & E94,\emph{NED}\\
14.83 & -2.80$\pm$0.02 & May 1986           & E94,\emph{NED}\\
14.73 & -2.73$\pm$0.01 & May 1986           & E94,\emph{NED}\\
14.64 & -2.76$\pm$0.02 & May 1986           & E94,\emph{NED}\\
14.53 & -2.78$\pm$0.02 & May 1986           & E94,\emph{NED}\\
14.38 & -2.75$\pm$0.02 & May 1986           & E94,\emph{NED}\\
14.38 & -2.74$\pm$0.02 & May 1986           & E94,\emph{NED}\\
14.26 & -2.79$\pm$0.02 & Apr. 1988          & E94,\emph{NED}\\
14.26 & -2.75$\pm$0.02 & Apr. 1988          & E94,\emph{NED}\\
17.38 & -6.36$\pm$0.17 & ...                & \emph{NED}    \\
17.48 & -6.51          & ...                & \emph{NED}    \\
17.50 & -5.47$\pm$0.01 & ...                & \emph{NED}    \\
17.66 & -6.25$\pm$0.05 & ...                & \emph{NED}    \\
17.99 & -7.00          & ...                & \emph{NED}    \\
18.08 & -6.67          & ...                & \emph{NED}    \\
18.10 & -6.80          & ...                & \emph{NED}    \\
18.10 & -6.40          & ...                & \emph{NED}    \\
18.16 & -7.26          & ...                & \emph{NED}

\enddata
\tablecomments{SED data of PG 1407+265 as plotted in Figures \ref{pg} and \ref{pg_ma}. The reference B96 refers to \citet{1996AJ....111.1431B}, H03: \citet{2003A&A...402...87H}, E94: \citet{1994ApJS...95....1E}, \emph{NED}: http://ned.ipac.caltech.edu/.}\label{pg_ta_sed}
\end{deluxetable}

\begin{deluxetable}{llllll}
\setlength{\tabcolsep}{0.07in}
\tablewidth{0pt}
\tabletypesize{\footnotesize}
\tablecaption{X-ray spectra of PG 1407+265}
\tablehead{
\colhead{Observation Date}&
\colhead{flux$^{\dag}$}&
\colhead{Photon index}&
\colhead{$f_{\rm2-10keV}^{\ddag}$}&
\colhead{Telescope}&
\colhead{Ref.}
}
\startdata

Jan. 17, 1981    & $f_{\rm1keV}=0.44^{+0.17}_{-0.05}$ $\mu$Jy                          & $\Gamma=2.2^{+1.7}_{-0.4}$     & 1.27 & EINSTEIN   & W87 \\
Jun. 19-20, 1987 & $f_{\rm2-10keV}=0.12\times10^{-11}$ erg/cm$^{2}$/s                  & ...                            & 1.2  & GINGA      & L97 \\
Jan. 19, 1992    & $f_{\rm1keV}=1.0^{+0.04}_{-0.04}$ $\mu$Jy                           & $\Gamma=2.61^{+0.05}_{-0.05}$  & 1.62 & ROSAT      & M95 \\
Jul. 02, 1993    & $f_{\rm2-10keV}=1.38^{+0.04}_{-0.04}\times10^{-12}$ erg/cm$^{2}$/s  & $\Gamma=2.05^{+0.05}_{-0.05}$  & 1.38 & ASCA       & G00 \\
Jan. 23, 2001    & $L_{\rm2-10keV}=8.85\times10^{45}$ erg/s                            & $\Gamma=2.32^{+0.02}_{-0.02}$  & 1.33 & XMM-Newton & G06 \\
Dec. 22, 2001    & $f_{\rm2-10keV}=0.8\times10^{-12}$ erg/cm$^{2}$/s                   & $\Gamma=2.24^{+0.01}_{-0.02}$  & 0.8  & XMM-Newton & F05

\enddata
\tablecomments{X-ray spectra of PG 1407+265. $^{\dag}$ flux presented in references. $^{\ddag}$ 2-10 keV integral flux (in unit of $10^{-12}$ erg/cm$^{2}$/s) calculated from the flux$^{\dag}$ and the $\Gamma$. The reference W87 refers to \citet{1987ApJ...323..243W}, L97: \citet{1997MNRAS.288..920L}, M95: \citet{1995ApJ...450..585M}, G00: \citet{2000ApJ...531...52G}, G06: \citet{2006MNRAS.365..960G}, F05: \citet{2005ApJ...633...61F}.}\label{pg_ta_x-ray}
\end{deluxetable}

\begin{deluxetable}{crrrrrrrrcccc}
\setlength{\tabcolsep}{0.07in}
\tablewidth{0pt}
\tabletypesize{\footnotesize}
\tablecaption{Parameters of PG 1407+265
}
\tablehead{
\colhead{Model}&
\colhead{$B$(Gs)}&
\colhead{$N_{0}$}&
\colhead{$R^{\dag}$}&
\colhead{$p_{1}$}&
\colhead{$p_{2}$}&
\colhead{$\gamma_{0}$}&
\colhead{$\gamma_{\rm min}$}&
\colhead{$\dot{m}$}&
\colhead{$\alpha$}&
\colhead{$\tau_{\rm r\varphi}$}&
\colhead{$L_{\rm jet}^{\ddag}$}&
\colhead{$L_{\rm disk}^{\ddag}$}
}
\startdata
1  & 29.4 & 1.7$\times10^{5}$ & 4.2 & 1.8 & 4.2 & 95    & 13.5 & 0.08 & 1.0 & $\alpha\sqrt{p_{\rm gas}p_{\rm tot}}$ & 11.3 & 68.1 \\
2  & 5.0  & 1.8$\times10^{3}$ & 4.2 & 1.8 & 4.2 & 23148 & 4.6  & 0.08 & 1.0 & $\alpha\sqrt{p_{\rm gas}p_{\rm tot}}$ & 0.33 & 68.1 \\
3  & ...  & ...               & ... & ... & ... & ...   & ...  & 0.3  & 0.5 & $\alpha p_{\rm tot}$                  & ...  & 144
\enddata
\tablecomments{$^{\dag}$ $R$ is the radius of emission region in unit of $10^{15}$ cm. $^{\ddag}$ $L_{\rm jet}$ and $L_{\rm disk}$ are in unit of 10$^{45}$ erg/s. The model parameters have the same meanings as those
in Table 1. Model 1 corresponds to the upper panel of Figure
\ref{pg}. Model 2 is for the lower panel of Figure \ref{pg}. Model
corresponds to the result in Figure \ref{pg_ma}.}\label{pg_ta}
\end{deluxetable}


\newpage


\begin{thebibliography}{99}
\bibitem[Abdo et al.(2010)]{2010ApJ...715..429A} Abdo, A.~A., Ackermann,
M., Ajello, M., et al.\ 2010, \apj, 715, 429


\bibitem[Ackermann et al.(2011)]{2011ApJ...743..171A} Ackermann, M.,
Ajello, M., Allafort, A., et al.\ 2011, \apj, 743, 171


\bibitem[Arp(1968)]{1968ApJ...152.1101A} Arp, H.\ 1968, \apj, 152, 1101


\bibitem[Barvainis et al.(1996)]{1996AJ....111.1431B} Barvainis, R.,
Lonsdale, C., \& Antonucci, R.\ 1996, \aj, 111, 1431


\bibitem[Berriman et al.(1990)]{1990ApJS...74..869B} Berriman, G., Schmidt,
G.~D., West, S.~C., \& Stockman, H.~S.\ 1990, \apjs, 74, 869


\bibitem[Blandford
\& Konigl(1979)]{1979ApJ...232...34B} Blandford, R.~D., \& Konigl, A.\ 1979, \apj, 232, 34


\bibitem[Blumenthal
\& Gould(1970)]{1970RvMP...42..237B} Blumenthal, G.~R., \& Gould, R.~J.\ 1970, Reviews of Modern Physics, 42, 237


\bibitem[Blundell
\& Beasley(1998)]{1998MNRAS.299..165B} Blundell, K.~M., \& Beasley, A.~J.\ 1998, \mnras, 299, 165


\bibitem[Blundell et al.(2003)]{2003ApJ...591L.103B} Blundell, K.~M.,
Beasley, A.~J., \& Bicknell, G.~V.\ 2003, \apjl, 591, L103


\bibitem[Broderick
\& Fender(2011)]{2011MNRAS.417..184B} Broderick, J.~W., \& Fender, R.~P.\ 2011, \mnras, 417, 184


\bibitem[Brunthaler et
al.(2005)]{2005A&A...435..497B} Brunthaler, A., Falcke, H., Bower, G.~C., et al.\ 2005, \aap, 435, 497


\bibitem[Brunthaler et
al.(2000)]{2000A&A...357L..45B} Brunthaler, A., Falcke, H., Bower, G.~C., et al.\ 2000, \aap, 357, L45


\bibitem[Cao(2009)]{2009MNRAS.394..207C} Cao, X.\ 2009, \mnras, 394, 207


\bibitem[Cao(2005)]{2005ApJ...619...86C} Cao, X.\ 2005, \apj, 619, 86


\bibitem[Celotti
\& Fabian(1993)]{1993MNRAS.264..228C} Celotti, A., \& Fabian, A.~C.\ 1993, \mnras, 264, 228


\bibitem[Celotti
\& Ghisellini(2008)]{2008MNRAS.385..283C} Celotti, A., \& Ghisellini, G.\ 2008, \mnras, 385, 283


\bibitem[Chen
\& Bai(2011)]{2011ApJ...735..108C} Chen, L., \& Bai, J.~M.\ 2011, \apj, 735, 108


\bibitem[Chen et al.(2010)]{2010RAA....10..707C} Chen, L., Bai, J.-M.,
Zhang, J.,
\& Liu, H.-T.\ 2010, Research in Astronomy and Astrophysics, 10, 707


\bibitem[Chen
\& Bai(2011)]{2011ScChG..54..183C} Chen, L., \& Bai, J.\ 2011, Science in China G: Physics and Astronomy, 54, 183


\bibitem[Cirasuolo et al.(2003)]{2003MNRAS.346..447C} Cirasuolo, M.,
Celotti, A., Magliocchetti, M., \& Danese, L.\ 2003, \mnras, 346, 447


\bibitem[Di Matteo(1998)]{1998MNRAS.299L..15D} Di Matteo, T.\ 1998, \mnras,
299, L15


\bibitem[Di Matteo et al.(1999)]{1999MNRAS.304..809D} Di Matteo, T.,
Celotti, A., \& Fabian, A.~C.\ 1999, \mnras, 304, 809


\bibitem[Elvis et al.(1994)]{1994ApJS...95....1E} Elvis, M., Wilkes, B.~J.,
McDowell, J.~C., et al.\ 1994, \apjs, 95, 1


\bibitem[Falcke et
al.(2004)]{2004A&A...414..895F} Falcke, H., K{\"o}rding, E., \& Markoff, S.\ 2004, \aap, 414, 895


\bibitem[Falcke et al.(1996)]{1996ApJ...471..106F} Falcke, H., Sherwood,
W., \& Patnaik, A.~R.\ 1996, \apj, 471, 106


\bibitem[Fang et al.(2005)]{2005ApJ...633...61F} Fang, T., Canizares,
C.~R., \& Marshall, H.~L.\ 2005, \apj, 633, 61


\bibitem[Fossati et al.(1998)]{1998MNRAS.299..433F} Fossati, G., Maraschi,
L., Celotti, A., Comastri, A., \& Ghisellini, G.\ 1998, \mnras, 299, 433


\bibitem[Galeev et al.(1979)]{1979ApJ...229..318G} Galeev, A.~A., Rosner,
R., \& Vaiana, G.~S.\ 1979, \apj, 229, 318


\bibitem[Gallo(2006)]{2006MNRAS.365..960G} Gallo, L.~C.\ 2006, \mnras, 365,
960


\bibitem[George et al.(2000)]{2000ApJ...531...52G} George, I.~M., Turner,
T.~J., Yaqoob, T., et al.\ 2000, \apj, 531, 52


\bibitem[Ghisellini et al.(1998)]{1998MNRAS.301..451G} Ghisellini, G.,
Celotti, A., Fossati, G., Maraschi, L.,
\& Comastri, A.\ 1998, \mnras, 301, 451


\bibitem[Ghisellini et
al.(1985)]{1985A&A...146..204G} Ghisellini, G., Maraschi, L., \& Treves, A.\ 1985, \aap, 146, 204


\bibitem[Ghisellini et al.(2010)]{2010MNRAS.402..497G} Ghisellini, G.,
Tavecchio, F., Foschini, L., et al.\ 2010, \mnras, 402, 497


\bibitem[Haardt
\& Maraschi(1991)]{1991ApJ...380L..51H} Haardt, F., \& Maraschi, L.\ 1991, \apjl, 380, L51


\bibitem[Haardt
\& Maraschi(1993)]{1993ApJ...413..507H} Haardt, F., \& Maraschi, L.\ 1993, \apj, 413, 507


\bibitem[Haas et
al.(2003)]{2003A&A...402...87H} Haas, M., Klaas, U., M{\"u}ller, S.~A.~H., et al.\ 2003, \aap, 402, 87


\bibitem[Ho
\& Peng(2001)]{2001ApJ...555..650H} Ho, L.~C., \& Peng, C.~Y.\ 2001, \apj, 555, 650


\bibitem[Hryniewicz et al.(2010)]{2010MNRAS.404.2028H} Hryniewicz, K.,
Czerny, B., Niko{\l}ajuk, M., \& Kuraszkiewicz, J.\ 2010, \mnras, 404, 2028


\bibitem[Hutchings
\& Campbell(1983)]{1983Natur.303..584H} Hutchings, J.~B., \& Campbell, B.\ 1983, \nat, 303, 584


\bibitem[Inoue
\& Takahara(1996)]{1996ApJ...463..555I} Inoue, S., \& Takahara, F.\ 1996, \apj, 463, 555


\bibitem[Ivezi{\'c} et al.(2002)]{2002AJ....124.2364I} Ivezi{\'c}, {\v Z}.,
Menou, K., Knapp, G.~R., et al.\ 2002, \aj, 124, 2364


\bibitem[Jang
\& Miller(1997)]{1997AJ....114..565J} Jang, M., \& Miller, H.~R.\ 1997, \aj, 114, 565


\bibitem[Jiang et al.(1998)]{1998ApJ...494..139J} Jiang, D.~R., Cao, X.,
\& Hong, X.\ 1998, \apj, 494, 139


\bibitem[Jim{\'e}nez-Bail{\'o}n et
al.(2005)]{2005A&A...435..449J} Jim{\'e}nez-Bail{\'o}n, E., Piconcelli, E., Guainazzi, M., et al.\ 2005, \aap, 435, 449


\bibitem[Kaastra
\& de Korte(1988)]{1988A&A...198...16K} Kaastra, J.~S., \& de Korte, P.~A.~J.\ 1988, \aap, 198, 16


\bibitem[Kardashev(1962)]{1962SvA.....6..317K} Kardashev, N.~S.\ 1962,
\sovast, 6, 317


\bibitem[Kawaguchi et al.(2001)]{2001ApJ...546..966K} Kawaguchi, T.,
Shimura, T., \& Mineshige, S.\ 2001, \apj, 546, 966


\bibitem[Kellermann et al.(1989)]{1989AJ.....98.1195K} Kellermann, K.~I.,
Sramek, R., Schmidt, M., Shaffer, D.~B., \& Green, R.\ 1989, \aj, 98, 1195


\bibitem[Kirk et
al.(1998)]{1998A&A...333..452K} Kirk, J.~G., Rieger, F.~M., \& Mastichiadis, A.\ 1998, \aap, 333, 452


\bibitem[Konigl(1981)]{1981ApJ...243..700K} Konigl, A.\ 1981, \apj, 243,
700


\bibitem[Kukula et al.(1998)]{1998MNRAS.297..366K} Kukula, M.~J., Dunlop,
J.~S., Hughes, D.~H., \& Rawlings, S.\ 1998, \mnras, 297, 366


\bibitem[Laor
\& Davis(2011)]{2011MNRAS.417..681L} Laor, A., \& Davis, S.~W.\ 2011, \mnras, 417, 681


\bibitem[Lawson
\& Turner(1997)]{1997MNRAS.288..920L} Lawson, A.~J., \& Turner, M.~J.~L.\ 1997, \mnras, 288, 920


\bibitem[Leipski et
al.(2006)]{2006A&A...455..161L} Leipski, C., Falcke, H., Bennert, N., H\"{u}ttemeister, S.\ 2006, \aap, 455, 161


\bibitem[Liu et al.(2002)]{2002ApJ...572L.173L} Liu, B.~F., Mineshige, S.,
\& Shibata, K.\ 2002, \apjl, 572, L173


\bibitem[Liu et al.(2008)]{2008ApJ...677..884L} Liu, H.~T., Bai, J.~M.,
Zhao, X.~H., \& Ma, L.\ 2008, \apj, 677, 884


\bibitem[Liu et al.(2006)]{2006ApJ...637..669L} Liu, Y., Jiang, D.~R.,
\& Gu, M.~F.\ 2006, \apj, 637, 669


\bibitem[Malkan
\& Sargent(1982)]{1982ApJ...254...22M} Malkan, M.~A., \& Sargent, W.~L.~W.\ 1982, \apj, 254, 22


\bibitem[McDowell et al.(1995)]{1995ApJ...450..585M} McDowell, J.~C.,
Canizares, C., Elvis, M., et al.\ 1995, \apj, 450, 585


\bibitem[McEnery et al.(2004)]{2004ASSL..304..361M} McEnery, J.~E.,
Moskalenko, I.~V.,
\& Ormes, J.~F.\ 2004, Cosmic Gamma-Ray Sources, 304, 361


\bibitem[Merloni
\& Fabian(2002)]{2002MNRAS.332..165M} Merloni, A., \& Fabian, A.~C.\ 2002, \mnras, 332, 165


\bibitem[Merloni
\& Fabian(2001)]{2001MNRAS.328..958M} Merloni, A., \& Fabian, A.~C.\ 2001, \mnras, 328, 958


\bibitem[Merloni et al.(2003)]{2003MNRAS.345.1057M} Merloni, A., Heinz, S.,
\& di Matteo, T.\ 2003, \mnras, 345, 1057


\bibitem[Miller et al.(1993)]{1993MNRAS.263..425M} Miller, P., Rawlings,
S., \& Saunders, R.\ 1993, \mnras, 263, 425


\bibitem[Nandra et al.(1997)]{1997ApJ...488L..91N} Nandra, K., George,
I.~M., Mushotzky, R.~F., Turner, T.~J.,
\& Yaqoob, T.\ 1997, \apjl, 488, L91


\bibitem[Neugebauer et al.(1979)]{1979ApJ...230...79N} Neugebauer, G., Oke,
J.~B., Becklin, E.~E., \& Matthews, K.\ 1979, \apj, 230, 79


\bibitem[Plotkin et al.(2010)]{2010ApJ...721..562P} Plotkin, R.~M.,
Anderson, S.~F., Brandt, W.~N., et al.\ 2010, \apj, 721, 562


\bibitem[Popovi{\'c} et al.(2003)]{2003ApJ...599..185P} Popovi{\'c}, L.~{\v
C}., Mediavilla, E.~G., Bon, E., Stani{\'c}, N.,
\& Kubi{\v c}ela, A.\ 2003, \apj, 599, 185


\bibitem[Rybicki
\& Lightman(1979)]{1979rpa..book.....R} Rybicki, G.~B., \& Lightman, A.~P.\ 1979, New York, Wiley-Interscience, 1979.~393 p.,


\bibitem[Sakimoto
\& Coroniti(1981)]{1981ApJ...247...19S} Sakimoto, P.~J., \& Coroniti, F.~V.\ 1981, \apj, 247, 19


\bibitem[Salvi et al.(2002)]{2002MNRAS.335..177S} Salvi, N.~J., Page,
M.~J., Stevens, J.~A., et al.\ 2002, \mnras, 335, 177


\bibitem[Schmidt
\& Green(1983)]{1983ApJ...269..352S} Schmidt, M., \& Green, R.~F.\ 1983, \apj, 269, 352


\bibitem[Schnopper et al.(1978)]{1978ApJ...222L..91S} Schnopper, H.~W.,
Delvaille, J.~P., Epstein, A., et al.\ 1978, \apjl, 222, L91


\bibitem[Shakura
\& Sunyaev(1973)]{1973A&A....24..337S} Shakura, N.~I., \& Sunyaev, R.~A.\ 1973, \aap, 24, 337


\bibitem[Shang et al.(2011)]{2011ApJS..196....2S} Shang, Z., Brotherton,
M.~S., Wills, B.~J., et al.\ 2011, \apjs, 196, 2


\bibitem[Shields(1978)]{1978Natur.272..706S} Shields, G.~A.\ 1978, \nat,
272, 706


\bibitem[Sikora et al.(1994)]{1994ApJ...421..153S} Sikora, M., Begelman,
M.~C., \& Rees, M.~J.\ 1994, \apj, 421, 153


\bibitem[Sikora et al.(2007)]{2007ApJ...658..815S} Sikora, M., Stawarz,
{\L}., \& Lasota, J.-P.\ 2007, \apj, 658, 815


\bibitem[Strateva et al.(2005)]{2005AJ....130..387S} Strateva, I.~V.,
Brandt, W.~N., Schneider, D.~P., Vanden Berk, D.~G.,
\& Vignali, C.\ 2005, \aj, 130, 387


\bibitem[Sun
\& Malkan(1989)]{1989ApJ...346...68S} Sun, W.-H., \& Malkan, M.~A.\ 1989, \apj, 346, 68


\bibitem[Svensson
\& Zdziarski(1994)]{1994ApJ...436..599S} Svensson, R., \& Zdziarski, A.~A.\ 1994, \apj, 436, 599


\bibitem[Taam
\& Lin(1984)]{1984ApJ...287..761T} Taam, R.~E., \& Lin, D.~N.~C.\ 1984, \apj, 287, 761


\bibitem[Tavecchio et al.(2001)]{2001ApJ...554..725T} Tavecchio, F.,
Maraschi, L., Pian, E., et al.\ 2001, \apj, 554, 725


\bibitem[Tavecchio et al.(1998)]{1998ApJ...509..608T} Tavecchio, F.,
Maraschi, L., \& Ghisellini, G.\ 1998, \apj, 509, 608


\bibitem[Ulvestad et al.(2005)]{2005ApJ...621..123U} Ulvestad, J.~S.,
Antonucci, R.~R.~J., \& Barvainis, R.\ 2005, \apj, 621, 123


\bibitem[Unger et al.(1987)]{1987MNRAS.228..521U} Unger, S.~W., Lawrence,
A., Wilson, A.~S., Elvis, M., \& Wright, A.~E.\ 1987, \mnras, 228, 521


\bibitem[Wang et al.(2006)]{2006ApJ...645..856W} Wang, T.-G., Zhou, H.-Y.,
Wang, J.-X., Lu, Y.-J., \& Lu, Y.\ 2006, \apj, 645, 856


\bibitem[White et al.(2000)]{2000ApJS..126..133W} White, R.~L., Becker,
R.~H., Gregg, M.~D., et al.\ 2000, \apjs, 126, 133


\bibitem[Wilkes
\& Elvis(1987)]{1987ApJ...323..243W} Wilkes, B.~J., \& Elvis, M.\ 1987, \apj, 323, 243


\bibitem[Zdziarski et al.(1999)]{1999MNRAS.303L..11Z} Zdziarski, A.~A.,
Lubi{\'n}ski, P., \& Smith, D.~A.\ 1999, \mnras, 303, L11



\end{thebibliography}
\end{document}